\documentclass[11pt]{article}

\pdfoutput=1

\usepackage{amsmath, amsthm, amssymb, bbm, setspace,bigints}
\usepackage[margin=1 in]{geometry}
\usepackage{caption}
\usepackage{subcaption}
\usepackage[toc,page]{appendix}
   
\usepackage[pdftex]{graphicx}
\usepackage{pdfpages}
\usepackage{epstopdf}
\usepackage{booktabs}
\usepackage{natbib}
\usepackage[colorlinks,citecolor=blue,urlcolor=blue,filecolor=blue]{hyperref}

\graphicspath{ {images/} }

\newcommand{\be} {\begin{eqnarray*}}
\newcommand{\ee} {\end{eqnarray*}}

\newcommand{\norm}[1]{\left\Vert#1\right\Vert}

\newcommand \bbR{\mathbb{R}}

\def\T{{ \mathrm{\scriptscriptstyle T} }}

\def\Gauss{{ \mathrm{N} }}

\newcommand{\ind}{\mathbbm{1}}

\def\m{\mathcal}
\def\mb{\mathbb}
\def\mr{\mathrm}

\newcommand{\2}{\\[2ex]}

\DeclareMathOperator{\diag}{\text{diag}}

\newtheorem{theorem}{Theorem}[section]

\newtheorem{proposition}[theorem]{Proposition}

\allowdisplaybreaks

\title{The Soft Multivariate Truncated Normal Distribution with Applications to Bayesian Constrained Estimation}
\date{}
\author{Allyson Souris \thanks{aesouris@stat.tamu.edu}
\\
\and
Anirban Bhattacharya \thanks{anirbanb@stat.tamu.edu}
\\
\and
Debdeep Pati \thanks{debdeep@stat.tamu.edu}
\\
\and
\multicolumn{1}{p{.7\textwidth}}{\centering\emph{Department of Statistics, Texas A\&M University, \\ College Station, TX}}
}

\begin{document}

\maketitle

\noindent {\bf Abstract.} We propose a new distribution, called the soft tMVN distribution, which provides a smooth approximation to the truncated multivariate normal (tMVN) distribution with linear constraints. An efficient blocked Gibbs sampler is developed to sample from the soft tMVN distribution in high dimensions. We provide theoretical support to the approximation capability of the soft tMVN and provide further empirical evidence thereof. The soft tMVN distribution can be used to approximate simulations from a multivariate truncated normal distribution with linear constraints, or itself as a prior in shape-constrained problems. \\[1ex]

\section{Introduction} 
The truncated multivariate normal (tMVN) distribution is routinely used as a prior distribution on model parameters in Bayesian shape-constrained regression. Structural constraints, such as monotonicity and/or convexity, are commonly induced by expanding the function in an appropriate basis where the constraints can be induced by imposing {\em linear constraints} on the coefficients; some examples of such a basis include piecewise linear functions \citep{dunson}, splines \citep{cai}, Bernstein polynomials \citep{wang}, and compactly supported basis functions \citep{maatouk, PhysRevC.99.055202}. Under a Gaussian or scale-mixture of Gaussian error distribution, the conditional posterior of the basis coefficients once again turns out to be truncated normal with linear constraints, necessitating sampling from a tMVN distribution for posterior inference. 

The problem of sampling from a tMVN distribution with linear constraints is also frequently encountered as a component of a larger Markov chain Monte Carlo (MCMC) algorithm to sample from the full conditional distribution of a constrained parameter vector. As a running example revisited on multiple occasions in this article, consider binary variables $y_i = \ind(z_i > 0)$, with $z = (z_1, \ldots, z_n)^\T$ a vector of latent Gaussian thresholds \citep{albert} and $w \in \mb R^q$ a vector of parameters/latent variables so that the joint distribution of $\theta = (z, w)$ follows a $\m N(\mu, \Sigma)$ distribution. It then immediately follows that the (conditional) posterior of $\theta \mid y, \mu, \Sigma$ follows a $\m N(\mu, \Sigma)$ distribution truncated to $\otimes_{i=1}^n \m C_i \, \otimes \mb R^q$, with $\m C_i = (0, \infty)$ or $(-\infty, 0)$ depending on whether $y_i = 1$ or $0$. Such latent Gaussian threshold models are ubiquitous in the analysis of binary and nominal data; examples include probit regression and its multivariate extensions \citep{albert, holmes, chib, obrien}, multinomial probit models \citep{mcculloch, zhang, johndrow}, tobit models \citep{tobin, polasek}, and binary Gaussian process (GP) classification models \citep{girolami} among others. 

In this article, we propose a new family of distributions called the soft tMVN distribution which replaces the hard constraints in a tMVN distribution with a smoothed or ``soft'' version using a logistic sigmoid function. The soft tMVN distribution admits a smooth log-concave density on the $d$-dimensional Euclidean space. Although the soft tMVN distribution is supported on the entire $d$-dimensional space, it can be made to increasingly concentrate most of its mass on a polyhedron determined by multiple linear inequality constraints, by tweaking a parameter. In fact, we show that the soft tMVN distribution approximates the corresponding tMVN distribution in total variation distance. 

Recognizing the soft tMVN distribution as the posterior distribution in a pseudo-logistic regression model, we develop an efficient blocked Gibbs sampler combining the Polya--Gamma data augmentation of \cite{polson2013} along with a structured multivariate normal sampler from \cite{bhattacharya}. In contrast, existing Gibbs samplers for a tMVN distribution sample the coordinates one-at-a-time from their respective full conditional univariate truncated normal distributions \citep{geweke, kotecha, damien, rodriguez}. The algorithm of Geweke is implemented in the \texttt{R} package \texttt{tmvtnorm} \citep{tmvtnorm}. While the Gibbs sampling procedure is entirely automated, it is well-recognized in a broader context that such one-at-a-time updates can lead to slow mixing, especially if the variables are highly correlated. We have additionally observed numerical instabilities in the \texttt{R} implementation for unconstrained dimensions exceeding 400.  While exact Hamiltonian Markov chain (HMC) algorithms to sample from tMVN \citep{pakman2014exact} are also popular, such algorithms are not suitable to sample from the soft tMVN, and leaf-frog steps with careful tuning are necessary to obtain good mixing.  There also exists accept-reject algorithms for the tMVN distribution that create exact samples from the distribution \citep{botev}. The algorithm of Botev is implemented in the  \texttt{R} package \texttt{TruncatedNormal} \citep{truncatednormal}. While exact samples are possible, when the acceptance probability becomes small, either the algorithm slows tremendously or approximate samples are produced. We typically saw small acceptance probabilities in the \texttt{R} implementation when the constrained dimension exceeded 200.   With such motivation, we propose to replace a tMVN distribution with its softened version inside a larger MCMC algorithm and use our sampling strategy for the soft tMVN distribution. In recent years, there has been several instances of such approximate MCMC (aMCMC) \citep{johndrow2015approximations} algorithms where the exact transition kernel of a Markov chain is replaced by an approximation thereof for computational ease. 

The soft tMVN distribution can also be used as a prior distribution in Bayesian shape-constrained regression problems as an alternative to the usual tMVN prior. Like the tMVN distribution, the soft tMVN distribution is conditionally conjugate for the mean in a Gaussian likelihood.  The soft tMVN can be viewed as a shrinkage prior which encourages shrinkage towards a linearly constrained region rather than being supported on the region. There is an interesting parallel between the soft tMVN distribution and global-local shrinkage priors used in sparse regression problems. The global-local priors replace the point mass (at zero) of the more traditional discrete mixture priors and rather encourage shrinkage towards the origin, with the motivation that a subset of the regression coefficients may have a small but non-negligible effect. Similarly, the soft tMVN prior favors the shape constraints while allowing for small departures. 

The rest of the article is organized as follows. In Section 2, we introduce the soft tMVN distribution as an approximation to the tMVN distribution and discuss its properties. In Section 3, we discuss various strategies to sample from a soft tMVN distribution, including a scalable Gibbs sampler suitable for high-dimensional situations. Section 4 contains a number of simulation examples to illustrate the efficacy of the proposed sampler as well as the approximation capability of the soft tMVN distribution. Section 5 contains an example of a Gibbs sampler for a shape-constrained model where the soft tMVN distribution is preferred as a prior over the tMVN distribution. We conclude with a discussion in Section 6.

\section{The soft tMVN distribution} 
Consider a tMVN distribution 
\begin{align}\label{eq:trunG}
\gamma(\theta) \propto e^{-\frac{1}{2} \, (\theta - \mu)^{\T} \Sigma^{-1} (\theta - \mu) } \, \ind_{\m C}(\theta),
\end{align}
where $\mu \in \mb R^d$, $\Sigma$ is a $d \times d$ positive definite matrix, and $\m C$ is described by $r \le d$ linear constraints, 
\begin{align*}
\m C = \bigg\{ \theta \in \mb R^d \!:\! s_i \,(a_i^{\T} \theta) \ge 0, \ i = 1, \ldots, r \bigg\},
\end{align*}
where $s_i \in \{1, -1\}$ denotes the sign of the $i$th inequality, and $a_i \in \mb R^d$. Without loss of generality, we assume the first $r$ coordinates to be constrained; this is mainly for notational convenience and can always be achieved by reordering the variables, if necessary. We also assume throughout that $\m C$ has positive $\mb R^d$-Lebesgue measure, so that the density $\gamma$ in \eqref{eq:trunG} is non-singular on $\mb R^d$. In the special case where $a_i = e_i$, the $i$th unit vector in $\mb R^d$ (with 1 at the $i$th coordinate and 0 elsewhere), the constraint set $\m C$ reduces to the form $\otimes_{i=1}^r \m C_i \otimes \mb R^q$ mentioned in the introduction. While this is an important motivating example, our approach works more generally for the type of constraints in the above display. 

Write, using the convention $0^0 = 1$,
\begin{align*}
\ind(\theta \in \m C) &= \prod_{i \in [r] \,:\, s_i = 1} \ind(a_i^{\T} \theta \ge 0) \,  \prod_{i \in [r] \,:\, s_i = -1} \ind(a_i^{\T} \theta < 0) 
\\
&= \prod_{i=1}^r \{\ind(a_i^{\T} \theta \ge 0) \}^{\ind(s_i = 1)} \, \{\ind(a_i^{\T} \theta < 0) \}^{\ind(s_i = -1)}. 
\end{align*}
Our main idea is to replace the indicator functions above with a smoothed or ``soft'' approximation. A rich class of approximations to the indicator function $\ind_{(0, \infty)}(\cdot)$ is provided by sigmoid functions, which are non-negative, monotone increasing, differentiable, and satisfy $\lim_{x \to \infty} \sigma(x) = 1$ and $\lim_{x \to -\infty} \sigma(x) = 0$. 
The cumulative distribution function of any absolutely continuous distribution on $\mb R$ which is symmetric about zero can be potentially used as a sigmoid function. Here, for reasons to be apparent shortly, we choose to use the logistic sigmoid function $\sigma(x) = 1/(1+e^{-x})$, which is the cdf of the logistic distribution. Specifically, define, for $\eta > 0$, 
\begin{align}\label{eq:err}
\sigma_{\eta}(x) = \frac{1}{1+e^{-\eta x}} = \frac{e^{\eta x}}{1 + e^{\eta x}}, \quad x \in \mb R, 
\end{align}
to be a scaled version of $\sigma(\cdot)$. The parameter $\eta$ controls the quality of the approximation, with larger values of $\eta$ providing increasingly better approximations to $\ind_{(0, \infty)}(\cdot)$. In fact, it is straightforward to see that 
\begin{align}\label{eq:approx_basic}
| \sigma_{\eta}(x) - \ind_{(0, \infty)}(x) | \le \frac{1}{1 + e^{\eta |x| }}, \quad x \in \mb R. 
\end{align}
It is also immediate that $(1 - \sigma_\eta(\cdot))$ is an approximation to $\ind_{(-\infty, 0)}(\cdot)$ with the same approximation error. 

We are now ready to describe our approximation scheme. Fixing some large $\eta$ and replacing the indicators by their respective sigmoidal approximations in \eqref{eq:trunG}, we obtain the approximation $\gamma_\eta$ to $\gamma$ as 
\begin{align}\label{eq:trunG_approx}
\gamma_\eta(\theta) \propto e^{-\frac{1}{2} \, (\theta - \mu)^{\T} \Sigma^{-1} (\theta - \mu) }  \ \prod_{i=1}^r \bigg(\frac{e^{\eta \,a_i^{\T}\theta}}{1 + e^{\eta \, a_i^{\T}\theta}} \bigg)^{\ind(s_i = 1)} \, \bigg(\frac{1}{1 + e^{\eta \, a_i^{\T}\theta}} \bigg)^{\ind(s_i = -1)},
\end{align}
for $\theta \in \mb R^d$. We refer to $\gamma_\eta$ as a soft tMVN distribution and generically denote it by $\m N_{\m C}^s(\mu, \Sigma)$. In the one-dimensional case, $\gamma_\eta(\theta) = \phi(\theta|\mu, \sigma)F(\theta)$ where $\phi(x|\mu, \theta)$ is the normal density with mean $\mu$ and variance $\sigma^2$ and $F(x)$ is the logistic distribution function. This is similar to a skew normal density, except in the skew normal density, $F(x)$ is the normal distribution function instead of the logistic distribution function \citep{Arellano}. It is immediate to note that $\gamma_\eta$ is a smooth (infinitely differentiable) density supported on $\mb R^d$. Further, a simple calculation shows that 
\begin{align*}
\nabla^2 \big( - \log \gamma_\eta(\theta) \big) = \Sigma^{-1} + \sum_{i=1}^r \frac{ \eta^2 \, e^{\eta \, a_i^{\T}\theta}}{(1 + e^{\eta \, a_i^{\T}\theta)^2} } \, a_i a_i^{\T} \succsim 0, 
\end{align*}
i.e., the Hessian matrix of the negative log density is positive definite. This implies that $\gamma_\eta$ is a log-concave density, which, in particular means $\gamma_\eta$ is unimodal. 
We collect these various observations about $\gamma_\eta$ in Proposition \ref{prop:approx}.
\begin{proposition}\label{prop:approx}
Let $\gamma$ and $\gamma_\eta$ be respectively defined as in \eqref{eq:trunG} and \eqref{eq:trunG_approx}. Then, $\gamma_\eta$ is an infinitely differentiable, unimodal, log-concave density on $\mb R^d$. Further, 
$$
\lim_{\eta \to \infty} \int_{\mb R^d} | \gamma_\eta(\theta) - \gamma(\theta) | \,d \theta = 0. 
$$
\end{proposition}
A proof is provided in the supplementary material. The last part of Proposition \ref{prop:approx} formalizes the intuition that $\gamma_\eta$ approximates $\gamma$ for large $\eta$ by showing that the $L_1$ distance between $\gamma_\eta$ and $\gamma$ converges to 0 as $\eta \to \infty$. An inspection of the proof for the $L_1$ approximation will reveal that we haven't used any particular feature of the logistic function and the argument can be extended to other sigmoid functions. 

\begin{figure}[h!]
\centering
\includegraphics[width = \columnwidth]{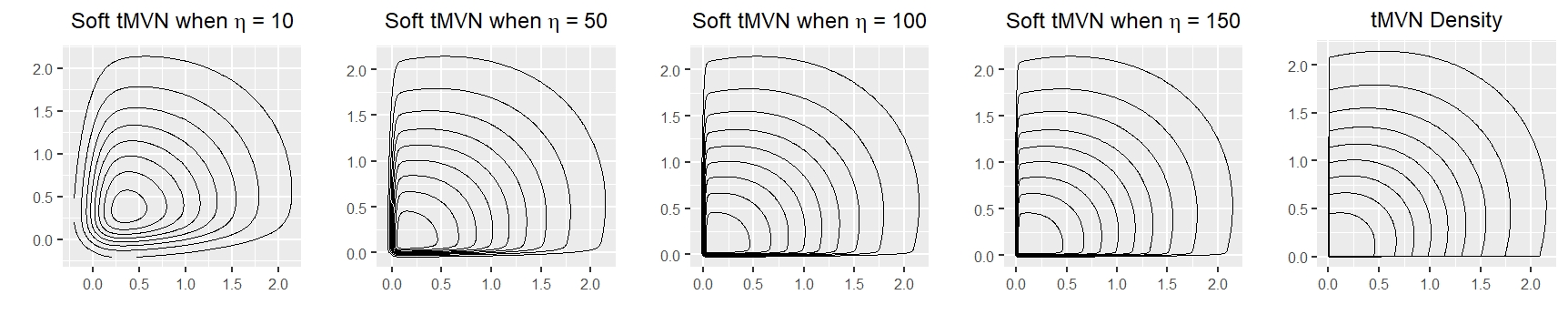}
\includegraphics[width = \columnwidth]{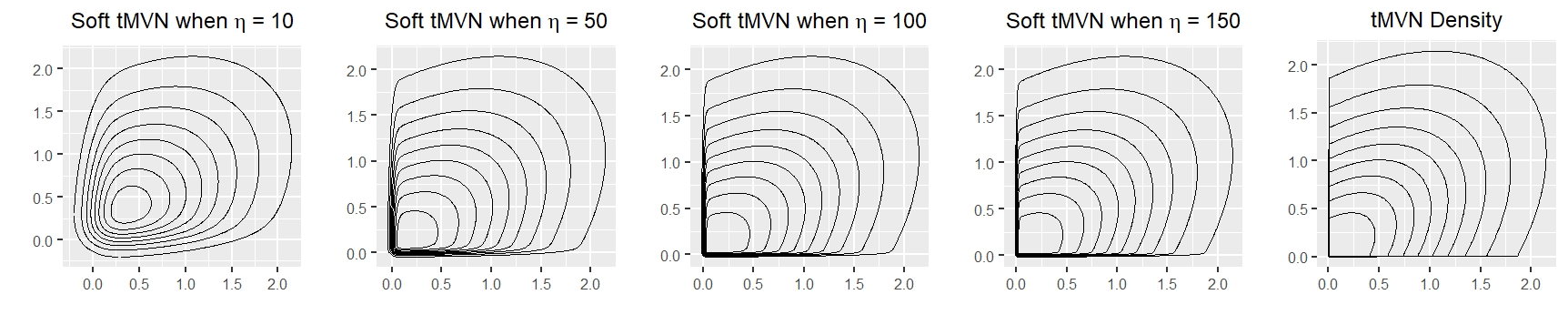}
\includegraphics[width = \columnwidth]{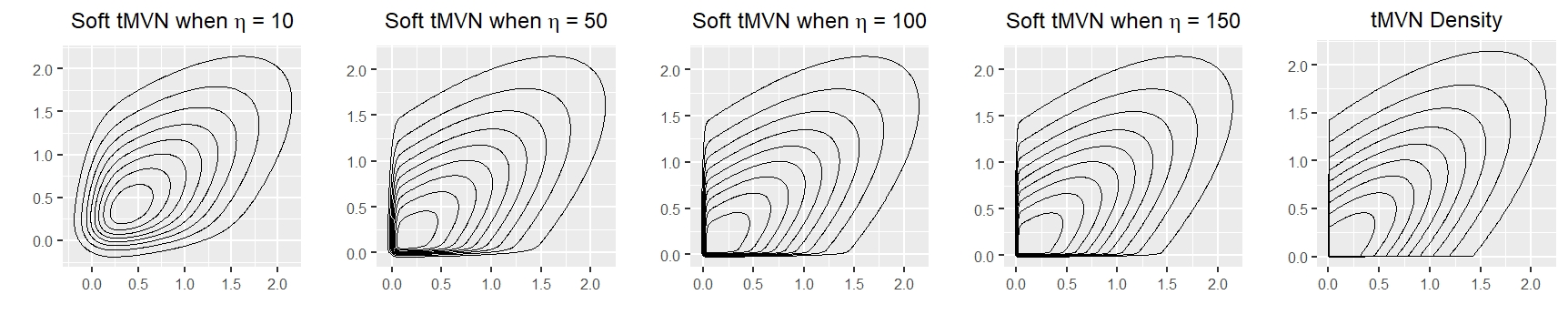}
\caption{{\em Contour plots of $\gamma$ and $\gamma_\eta$ for $\eta = 10, 50, 100$, and $150$, where $\gamma$ as in \eqref{eq:biv_ex} is a standard bivariate normal distribution with correlation $\rho$, truncated to the positive orthant. The rows from top to bottom correspond to $\rho = 0.25, 0.50,$ and $0.75$ respectively. } }
\label{fig:compTN}
\end{figure}

The $L_1$ approximation result implies that although $\gamma_\eta$ has a non-zero density at all points in $\mb R^d$, the effective support is the region $\m C$ for large values of $\eta$, and a random draw from $\gamma_\eta$ will fall inside $\m C$ with overwhelmingly large probability. This is because 
{\small
\begin{align*}
\gamma_\eta(\theta \not\in \m C) &= 1 - \gamma_\eta(\theta \in \m C) 
\\
&= \gamma(\theta \in \m C) - \gamma_\eta(\theta \in \m C) 
\\
&\leq \int_{\mb R^d} | \gamma_\eta(\theta) - \gamma(\theta) | \,d \theta,
\end{align*} 
}
so using Proposition \ref{prop:approx}, the probability of $\theta$ falling outside of the region $\m C$ approaches zero as $\eta$ approaches infinity.  To obtain a more quantitative feel for how the approximation gets better with increasing $\eta$, we set $\gamma$ to be a standard bivariate normal distribution truncated to the first orthant, 
\begin{align}\label{eq:biv_ex}
\gamma(\theta) \propto e^{-\theta^{\T} \Sigma^{-1} \theta} \, \ind_{(0, \infty)}(\theta_1) \, \ind_{(0, \infty)}(\theta_2), \quad 
\Sigma = 
\begin{pmatrix} 
1 & \rho \\
\rho & 1
\end{pmatrix}.
\end{align}
Figure \ref{fig:compTN} shows contour plots of $\gamma$ (last column) along with those for $\gamma_\eta$ for various values of $\eta$, with $\eta$ increasing from left to right. Each row corresponds to a different value of $\rho$. It is evident that the approximation quickly improves as $\eta$ increases, and stabilizes around $\eta = 100$. We later show in simulations involving substantial higher dimensions that $\gamma_\eta$ with $\eta = 100$ continues to provide a reasonable approximation to the corresponding tMVN distribution $\gamma$. 

The accurate approximation of the soft tMVN has two important consequences in our opinion. First, for any of the examples discussed in the introduction which require a sample from a tMVN within an MCMC algorithm, a sample from a tMVN can be replaced with a sample from the corresponding soft tMVN distribution; we discuss efficient strategies to sample the soft tMVN distribution in the next section. Second, the soft tMVN distribution can itself be used as a prior distribution for constrained parameters. As a prior, the soft tMVN replaces the hard constraints imposed by the tMVN with soft constraints, encouraging shrinkage towards the constrained region $\m C$. Indeed, the soft tMVN distribution can be considered a global shrinkage prior \citep{polson2010} which shrinks vectors towards a pre-specified constrained region. 

The tMVN prior is conditionally conjugate for a Gaussian likelihood and the soft tMVN prior naturally inherits this conditional conjugacy. Suppose $Y \mid \theta, \sigma^2 \sim \m N(\Phi \theta, \sigma^2 I_n)$ and $\theta \sim \m N_{\m C}^s(\mu, \Sigma)$ is assigned a soft tMVN prior. Then, 
$$
\theta \mid Y, \sigma^2, \mu, \Sigma \sim \m N_{\m C}^s\big( (\Phi^{\T} \Phi/\sigma^2 + \Sigma^{-1})^{-1} \Phi^{\T} Y, \, (\Phi^{\T} \Phi/\sigma^2 + \Sigma^{-1})^{-1} \big). 
$$
The conditional conjugacy allows one to fit a conditionally Gaussian model with a soft tMVN prior using standard Gibbs sampling algorithms, provided one can efficiently sample from a soft tMVN distribution. We provide a detailed exposition in Section \ref{sec:cons}, with a specific application of the soft tMVN distribution as a prior in Bayesian monotone single-index models. 

\section{Sampling from the soft tMVN distribution}

\subsection{Gibbs sampler in high-dimensions}
In this subsection, we propose a scalable data-augmentation blocked-Gibbs sampler to sample from a soft tMVN distribution. The proposed Gibbs sampler updates the entire $\theta$ vector in a block, unlike one-at-a-time updates for Gibbs samplers for tMVNs. 

Apart from log-concavity, the other nice feature behind our choice of the logistic sigmoid function is that $\gamma_\eta$ can be recognized as the posterior distribution of a vector of regression parameters in a logistic regression model. To see this, consider the setup of a logistic regression model with binary response $t_i \in \{0, 1\}$ and vector of predictors $W_i \in \mb R^d$ for $i = 1, \ldots r$, 
$$
\mbox{Pr}(t_i = 1 \mid \theta, W_i) = \frac{e^{W_i^{\T} \theta}}{1+e^{W_i^{\T}\theta}}. 
$$
Assuming a $\m N(\mu, \Sigma)$ prior on the vector of regression coefficients $\theta$, the posterior distribution of $\theta \mid t, W, \mu, \Sigma$ is given by 
$$
e^{-\frac{1}{2} \, (\theta - \mu)^{\T} \Sigma^{-1} (\theta - \mu) }  \ \prod_{i=1}^r \bigg(\frac{e^{W_i^{\T}\theta}}{1 + e^{W_i^{\T}\theta}} \bigg)^{t_i} \, \bigg(\frac{1}{1 + e^{W_i^{\T}\theta}} \bigg)^{(1-t_i)}. 
$$
If we now set $t_i = \ind(s_i = 1)$ and $W_i = \eta \, a_i$, then the above density is identical to $\gamma_\eta$. The number of constraints $r$ plays the role of the sample size, and the ambient dimension $d \ge r$ indicates the number of the regression parameters in this pseudo-logistic model. Thus, sampling from $\gamma_\eta$ is equivalent to sampling from the conditional posterior of regression parameters in a high-dimensional logistic regression model, which can be conveniently carried out using the Polya--Gamma data augmentation scheme of \cite{polson2013}. The Polya--Gamma scheme introduces $r$ auxiliary variables $\omega_1, \ldots, \omega_r$ and performs Gibbs sampling by alternatively sampling from $\omega \mid \theta, t$ and $\theta \mid \omega, t$ as follows: \\[2ex]
(i) Sample $\omega_i \mid \theta, t \sim \mbox{PG}(1,W_i^{\T}\theta)$ independently for $i = 1, \ldots, r$, \\
(ii) Sample $\theta \mid \omega, t \sim \m N_d(\mu_\omega, \Sigma_\omega)$, with
\begin{align}\label{eq:structured_G}
\Sigma_\omega = (W^{\T} \Omega W + \Sigma^{-1})^{-1}, \quad \mu_{\omega} = \Sigma_{\omega} (W^{\T} \kappa + \Sigma^{-1} \mu),
\end{align}
where $W \in \mb R^{r \times d}$ with $i$th row $W_i^{\T}$, \,$t = (t_1, \ldots, t_r)^{\T}$, \, $\kappa = (t-1/2)$, and $\Omega = \mbox{diag}(\omega_1, \ldots, \omega_r)$.  
\\[2ex]
In (i), PG denotes a Polya--Gamma distribution which can be sampled using the \texttt{Bayeslogit} package in \texttt{R} \citep{polson2013}. 
Note that the entire $\theta$ vector is sampled in a block in step (ii). The worst-case complexity of sampling from the multivariate Gaussian distribution in \eqref{eq:structured_G} is $O(d^3)$. However, exploiting the structure of $\mu_\omega$ and $\Sigma_\omega$, a sample from $\m N(\mu_\omega, \Sigma_\omega)$ can be obtained with significantly less cost using a recent algorithm in \cite{bhattacharya} provided $d \gg r$ and a $\m N(0, \Sigma)$ variate can be cheaply sampled. 

Define $\Phi = \Omega^{1/2} W$ and $\alpha = \Omega^{-1/2} \kappa$. Then, a sample from (ii) is obtained by first sampling 
\begin{align}\label{eq:bar_theta}
\bar{\theta} \sim \m N( (\Phi^{\T}\Phi + \Sigma^{-1})^{-1} \Phi^{\T} \alpha, \, (\Phi^{\T}\Phi + \Sigma^{-1})^{-1}),
\end{align}
and setting 
\begin{align}\label{eq:bar_mu}
\theta = \bar{\mu} + \bar{\theta}, \quad \bar{\mu} = (\Phi^{\T}\Phi + \Sigma^{-1})^{-1} \Sigma^{-1} \mu. 
\end{align}
First, by the Sherman--Woodbury--Morrison formula, 
$$
(\Phi^{\T} \Phi + \Sigma^{-1})^{-1} = \Sigma - \Sigma \Phi^{\T}(\Phi \Sigma \Phi^{\T} + I_r)^{-1}\Phi \Sigma.
$$
Thus, 
\begin{align}\label{eq:bar_mu1}
\bar{\mu} = \mu - \Sigma \Phi^{\T}(\Phi \Sigma \Phi^{\T} + I_r)^{-1}\Phi \mu,
\end{align}
which only requires solving a $r \times r$ system. 

Sampling $\bar{\theta}$ in \eqref{eq:bar_theta} can be efficiently carried out by adapting the algorithm of \cite{bhattacharya} to the present setting. The steps are: \\[1ex]
(a) Sample $u \sim \m N(0, \Sigma)$ and $\delta \sim \m N(0, \mr I_r)$. \\
(b) Set $v = \Phi u + \delta$. \\
(c) Solve $(\Phi \Sigma \Phi^{\T} + \mr I_r) w = (\alpha - v)$. \\
(d) Set $\bar{\theta} = u + \Sigma \Phi^{\T} w$. 
\2
It follows from \cite{bhattacharya} that $\bar{\theta}$ obtained in step (d) has the desired Gaussian distribution. Barring the sampling of $u$ in step (a), the remaining steps have a combined complexity of $O(r^2 d)$, which can be significantly smaller than $d^3$ when $d \gg r$.  If $\Sigma$ is a diagonal matrix, $u$ can be trivially sampled with $O(d)$ cost. Even for non-diagonal $\Sigma$, it is often possible to exploit its structure to cheaply sample from $\m N(0, \Sigma)$.  For example, in the probit and multivariate probit regression context, $\Sigma$ assumes the form (see Section \ref{sect:probit}), 
$$\Sigma = \begin{pmatrix} \mathrm{I}_N + H L H^{\T} & HL \\ L H^{\T} & L \end{pmatrix},$$
where $L$ is a $q\times q$ diagonal matrix and $H$ is an $N \times q$ (possibly dense) matrix. A sample $u$ from $\m N(0, \Sigma)$ is then obtained by \\[1ex]
(i) Sample $z \sim \m N(0, \mathrm{I}_N)$ and $u_2 \sim \m N(0, L)$ independently. \\
(ii) Set $u_1 = H u_2 + z$ and $u = (u_1^{\T}, u_2^{\T})^{\T}$. 
Since $u$ is a linear transformation of $(z, u_2)$ which is jointly Gaussian, $u$ also has a joint Gaussian distribution. Calculating the covariance matrix of $u$ then immediately shows that $u \sim \m N(0, \Sigma)$. Since $L$ is diagonal, $u_2$ can be sampled in $O(q)$ steps, and the matrix multiplication costs $O(N q^2)$, so that the overall cost is $O(N q^2)$. 

\subsection{Other strategies}
In moderate dimensions, it is possible to use a Metropolis (Gaussian) random walk and its various extensions to sample from a soft tMVN distribution. In particular, given that the soft tMVN distribution can be recognized as the posterior distribution in a model with a Gaussian prior,  elliptical slice sampling \citep{murray2010elliptical} is a viable option. 

There is substantial literature on sampling from log-concave distributions using variants of the Metropolis algorithm with strong theoretical guarantees \citep{frieze1994, frieze1999, lovasz2006fast, lovasz2006simulated, belloni}.  More recently, \cite{dalalyan} and \cite{durmus} provided non-asymptotic bounds on the rate of convergence of unadjusted Langevin Monte Carlo (LMC) algorithms for log-concave target densities. Assuming the target density is proportional to $e^{-f(\theta)}$ for some convex function $f$, the successive iterates of a first-order LMC algorithm takes the form 
\begin{align*}
\theta_{k+1} = \theta_k - h \nabla f(\theta_k) + \sqrt{2h} \, \xi_{k+1}, \quad k = 0, 1, \ldots, 
\end{align*}
where the $\{\xi_k\}$s are independent $\m N(0, \mathrm{I})$ variates and $h > 0$ is a step-size parameter. Clearly, $\{\theta_k\}_{k=0,1, \ldots}$ forms a discrete-time Markov chain and the results in \cite{dalalyan} and \cite{durmus} characterize the rate at which the distribution of $\theta_k$ converges to the target density in total variation distance. Aside from the non-asymptotic bounds, another key message from their results is that the typical Metropolis adjustment as in Metropolis adjusted Langevin (MALA) \citep{roberts} is not required for log-concave targets. \cite{dalalyan} also provides a second-order version of the LMC algorithm called LMCO which can incorporate the Hessian $\nabla^2 f$. Since both $\nabla (- \log \gamma_\eta)$ and $\nabla^2 ( - \log \gamma_\eta)$ are analytically tractable, it is possible to use both the LMC and LMCO algorithms to sample from $\gamma_\eta$. 

Other than MCMC, another possible strategy to sample from $\gamma_\eta$ is to use a multivariate generalization of the adaptive rejection sampling (ARS) \citep{gilks}. 

\begin{figure}[h!]
\centering
\includegraphics[width = 0.45\columnwidth]{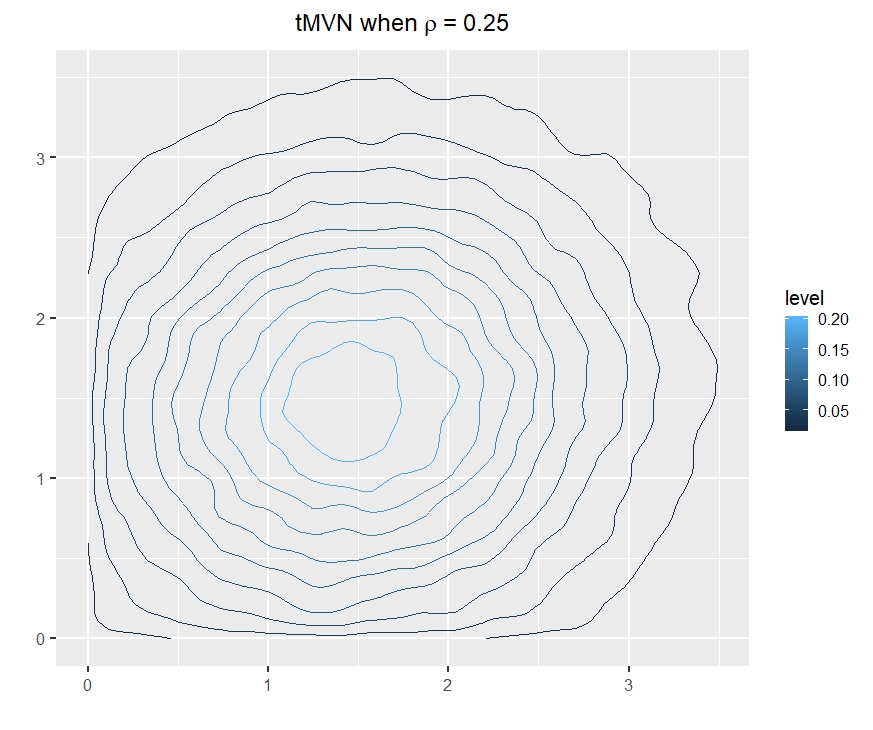}
\includegraphics[width = 0.45\columnwidth]{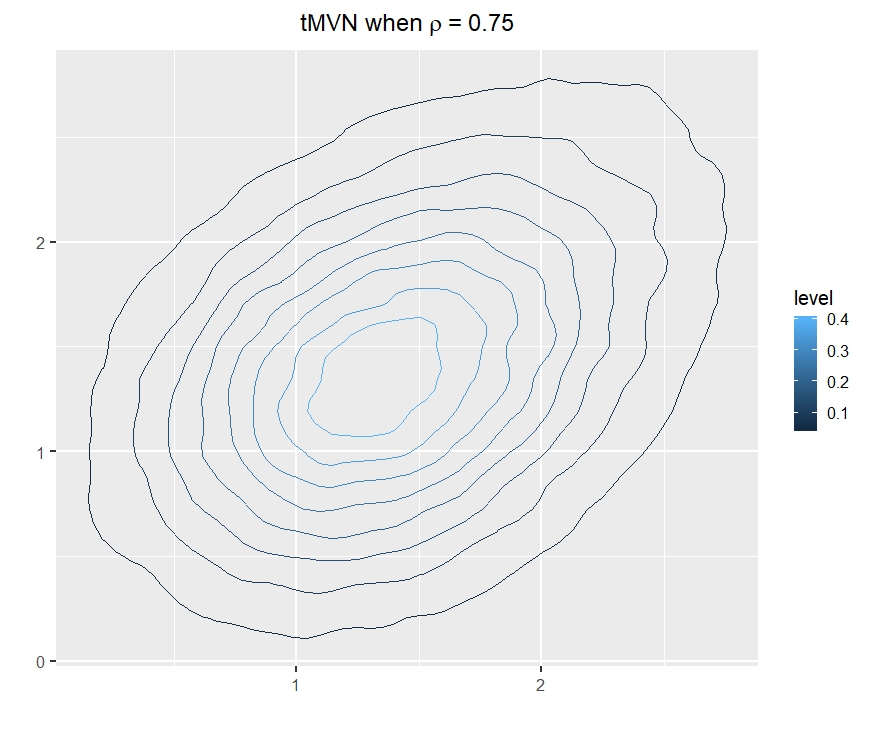}\\
\includegraphics[width = 0.45\columnwidth]{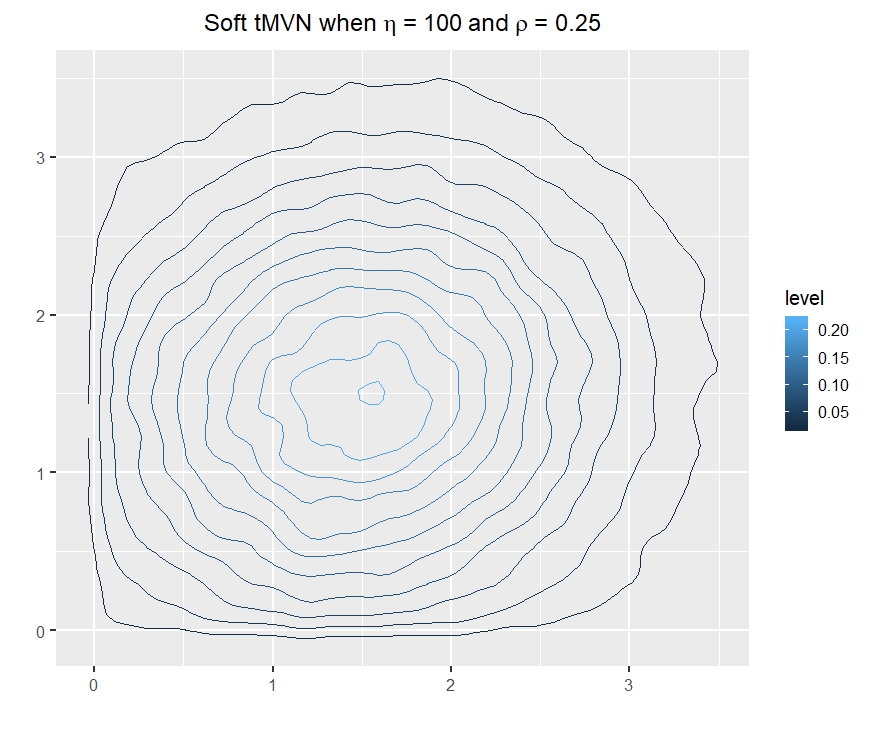}
\includegraphics[width = 0.45\columnwidth]{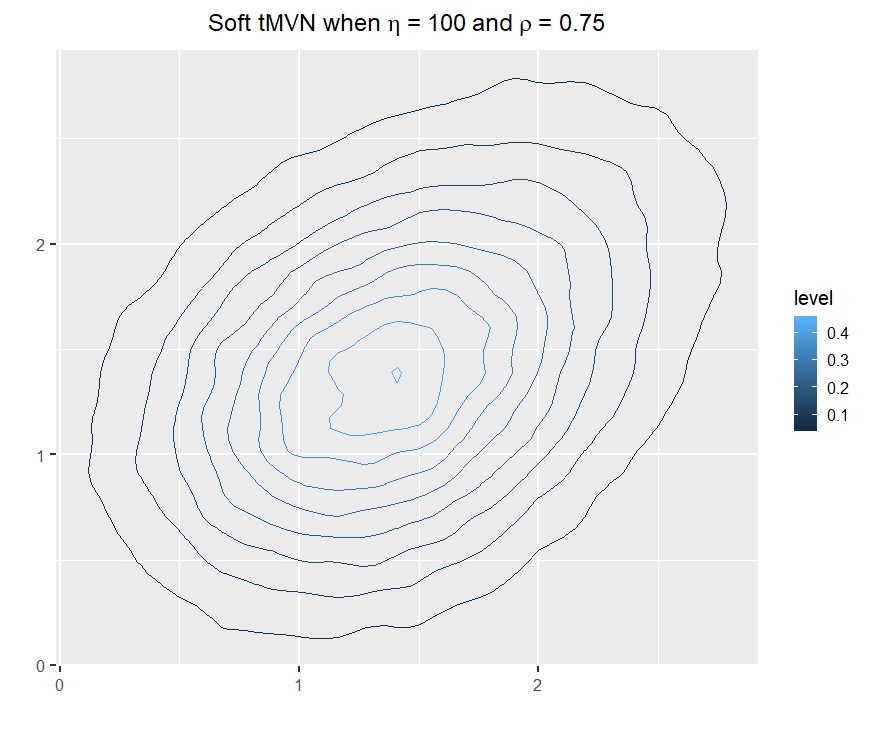}
\caption{{\em The top panel shows contour plots of a bivariate marginal of a 50-dimensional tMVN distribution with an equicorrelation covariance structure obtained using Botev's rejection sampler; the left and right figures correspond to the correlation parameter $\rho = 0.25$ and $0.75$ respectively. The bottom panel shows the same for the corresponding soft tMVN distribution with $\eta = 100$, which continues to provide a good approximation.} }
\label{fig:con_bi}
\end{figure}

\section{Simulations}

In this section, we conduct a number of simulations to empirically illustrate that the soft tMVN distribution continues to provide an accurate approximation to the tMVN distribution in high-dimensional situations. These simulations also demonstrate the scalability of the proposed Gibbs sampler. 

To begin with, we first justify our continued use of $\eta = 100$ in higher dimensions. In Figure \ref{fig:compTN}, we had provided the contour plots of a bivariate tMVN distribution and its soft tMVN approximation with $\eta = 100$. As an obvious extension, we now consider the bivariate marginal of $(\theta_1, \theta_2)$, where $\theta \in \bbR^{50}$ is drawn from a multivariate normal distribution with mean $\mu = 0$ and with a compound symmetry covariance structure, $\Sigma = (1-\rho) I_{50} + \rho 1_{50}1_{50}^{\T}$, truncated to the positive orthant. We consider two choices of $\rho$, namely $\rho = 0.25$ and $0.75$, and provide the contour plots for $\m N_{\m C}(\mu, \Sigma)$ and $\m N_{\m C}^s(\mu, \Sigma)$ in the top and bottom panels of Figure \ref{fig:con_bi} respectively. The contour plots were drawn by collecting $150,000$ samples from the $\m N_{\m C}(\mu, \Sigma)$ and $\m N_{\m C}^s(\mu, \Sigma)$ distributions, and then retaining the first two coordinates in each case to obtain samples from the bivariate marginal. Specifically, we used the rejection sampler of \cite{botev} implemented in the \texttt{R} package \texttt{TruncatedNormal} \citep{truncatednormal} to draw samples from a tMVN distribution and used our data augmentation Gibbs sampler to sample from the soft tMVN distribution. The figure shows that $\eta = 100$ remains a reasonable choice in higher dimensions, and we henceforth fix $\eta = 100$ throughout. The figure also shows that the contours between between the two distributions are comparable with the soft tMVN having a slightly larger peak.

Next, we provide some numerical summaries in two different settings. Due to the inherent difficulty of comparing two high-dimensional distributions, we will compare the marginal densities. Specifically, given densities $f$ and $g$ on $\mb R^d$ with finite mean, we consider two different measures to compare them. The first one uses the 1\textsuperscript{st} Wasserstein ($\mbox{W}_1$) distance between two distributions, $W_1(f,g)$ \citep{OptimalTransport}. The $\mbox{W}_1$ distance is defined as
$$
W_1(f, g) = \inf_{(U, V) \in \mathcal{C}_{f, g}} \, E\|U-V\|
$$
where $\mathcal{C}_{f, g}$ is the collection of all couplings between $f$ and $g$, i.e., pair of random variables $(U, V)$ with $U \sim f$ and $V \sim g$. Our first comparison metric is an average $\mbox{W}_1$ distance between the marginals, 
\begin{align}\label{eq:D}
D :\,= \frac{1}{d} \sum_{i=1}^d \mbox{W}_1(f_i, g_i),
\end{align}
where $f_i$ denotes the $i$th marginal density of $f$. We used the R package \texttt{transport} to compute the average $W_1$ distance between $\gamma$ and $\gamma_\eta$, which to our convenience only requires samples from the two densities in questions. We note here that an analytic calculation is out of question since the marginal densities of both $\gamma$ and $\gamma_\eta$ lack closed-form expressions.

Our second measure is an average squared $L_2$ distance between the mean vectors for the two densities, 
\begin{align}\label{eq:xi}
\xi :\, = \frac{\|\mu_f - \mu_g\|^2}{d}, 
\end{align}
with $\mu_f = \int_{\mb R^d} x f(x) dx$. 

We compute $D$ and $\xi$ between $\gamma$ and $\gamma_\eta$ for two different covariance structures in $\Sigma$. Due to the lack of analytic expressions for the marginals for non-diagonal $\Sigma$, we resort to simulations to approximate $D$ and $\xi$.  The highest dimension $d$ used in our simulations is $d = 600$; while our sampler can be scaled beyond this, the rejection sampler starts producing warning messages due to incurring small acceptance probabilities. The code for sampling from the soft tMVN distribution with both covariance structures is located at \url{https://github.com/aesouris/softTMVN}.

\subsection{Probit-Gaussian Process Example}

For our first example, we consider $\theta \sim \m N_{n}(0, \Sigma)\ind_{\m{C}}(\theta)$ where the covariance matrix $\Sigma$ is formed from the M{\'a}tern kernel \citep{rasmussen2004gaussian} and $\m{C} = \m{C}_1 \otimes \m{C}_2 \otimes \cdots \otimes \m{C}_n$ where $\m{C}_i$ is either $(-\infty, 0)$ or $(0, \infty)$ for $i = 1, \ldots, n$. This structure is motivated by a binary Gaussian process (GP) classification model. Suppose $Y_i \in \{0,1\}$ is a binary response at locations $s_i$ modeled as $Y_i = \ind\{Z(s_i) > 0\}$ for $i = 1, \ldots, n$, where $Z$ is a continuous latent threshold function. In GP classification, $Z$ is assigned a mean-zero Gaussian process prior
$Z \sim GP(0, K_n)$, with $[K_n]_{ij} = K(s_i, s_j)$ and $K$ a positive definite kernel. Here, we take $K$ to be a M{\'a}tern kernel. 
Letting $Z = [Z(s_1), \ldots, Z(s_n)]^{\T}$, the conditional distribution of $Z \mid Y$ follows the above $\m N_{n}(0, K_n) \ind_{\m{C}}(Z)$ where $\m{C}_i = (-\infty, 0)$ if $Y_i = 0$ and $\m{C}_i = (0, \infty)$ if $Y_i = 1$. 

For the simulation, set $n = \{100, 200\}$. Let $s_i = i$ for $i = 1, \ldots, n$. We randomly sample $\ell_1$ from $\{10,\ldots, n/2\}$ and $\ell_2$ from $\{n/2+1, \ldots, n-10\}$ and let $Y_1, \ldots, Y_{\ell_1} = 1$, $Y_{\ell_1+1}, \ldots, Y_{\ell_2} = 0$, and $Y_{\ell_2 + 1}, \ldots, Y_n = 1$. This is simply to mimic the situation when the true latent function $Z$ takes positive values on $[0, a]$, negative values on $[a, b]$ and positive values again on $[b, \infty]$ for some $0 < a < b$. We set the smoothness parameter for the M{\'a}tern kernel at 3/5 and the scale parameter at 1. We then proceed to draw 5000 samples from the tMVN, $\m N_{n}(0, \Sigma)\ind_{\m{C}}(\theta)$, using Botev's rejection sampler and 5000 samples from the soft tMVN, $\m N_{n}^s(0, \Sigma)\ind_{\m{C}}(\theta)$, using our Gibbs sampler. The 5000 samples were collected for our method after discarding 1000 initial samples as burn-in and collecting every 100th sample to thin the chain. There is high autocorrelation in the chain, so the large thinning parameter is necessary, but this is an effecient sampler, so we are not worried about the extra sampling. 

\begin{figure}[h]
\centering
\includegraphics[width = 1.1\columnwidth]{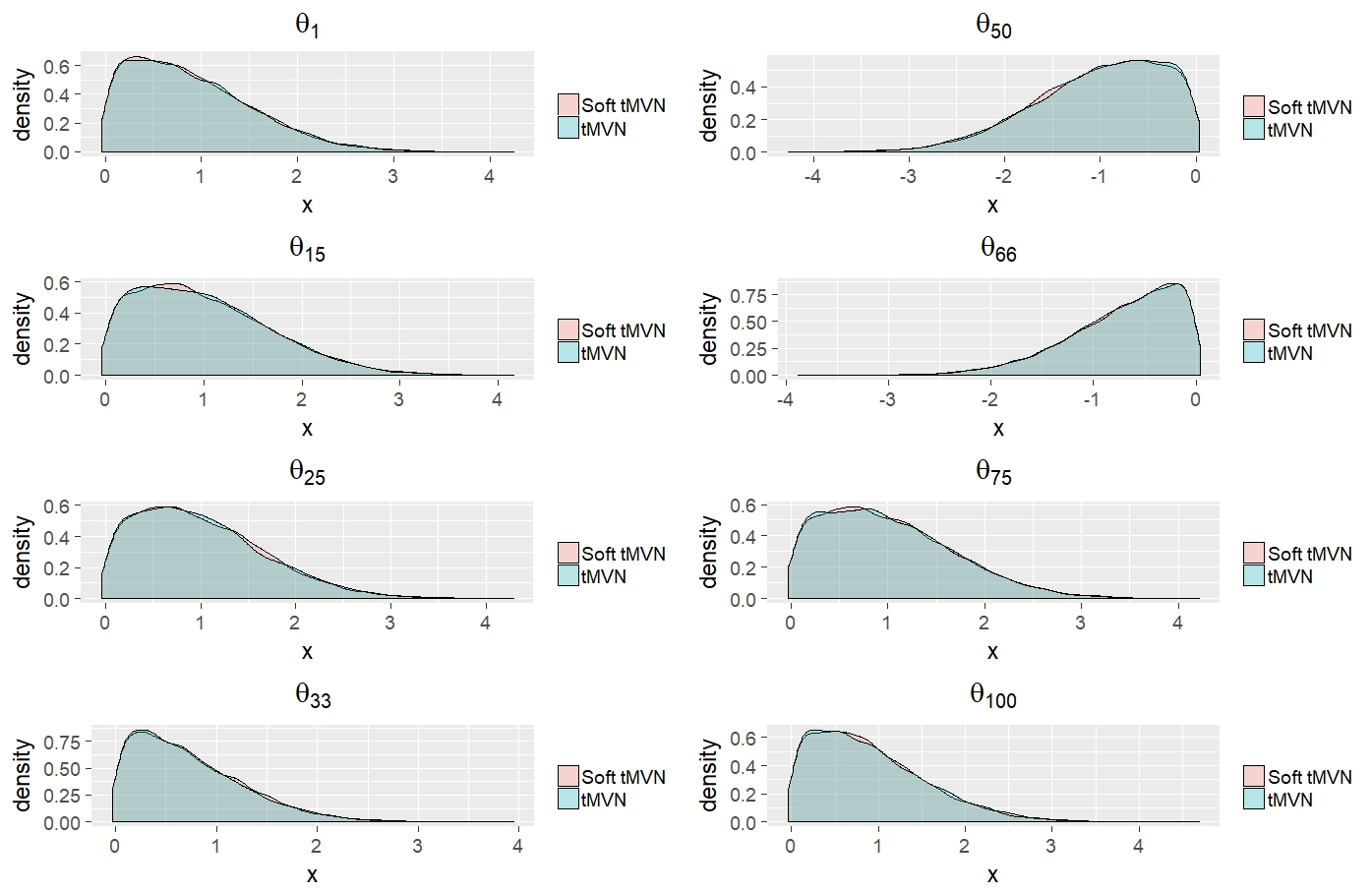}
\caption{{\em Overlapping density plot for the Probit-Gaussian Process simulation when $n = 100$. Blue denotes tMVN using Botev's rejection sampler and pink denotes the soft tMVN distribution. The density plots are obtained using 5000 independent samples from each distribution. 
}}
\label{fig:DP3}
\end{figure}

\begin{figure}[h]
\centering
\includegraphics[width = 1.1\columnwidth]{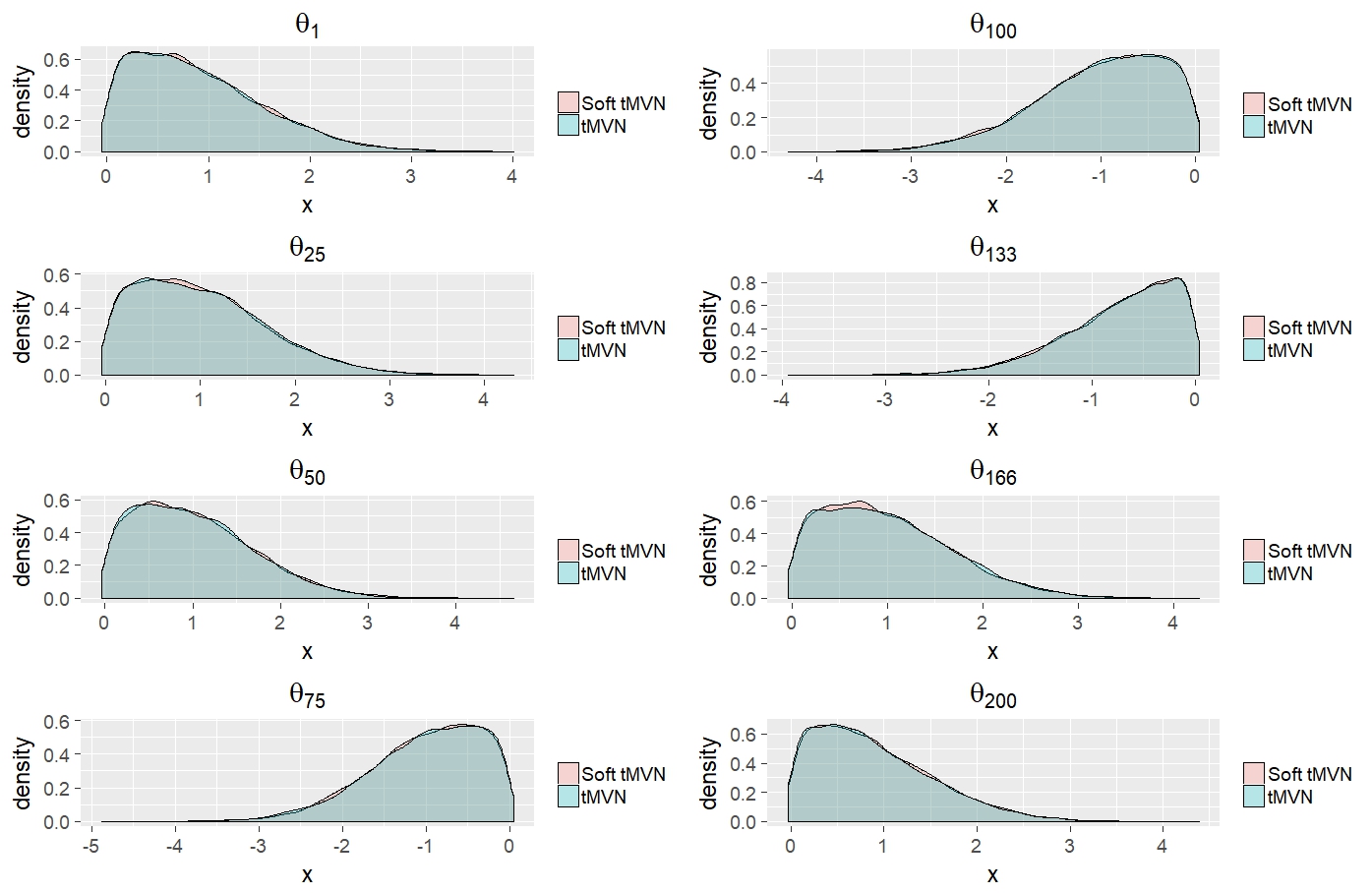}
\caption{{\em Overlapping density plot for the Probit-Gaussian Process simulation when $n = 200$. Blue denotes tMVN and pink denotes the soft tMVN distribution. The density plots are obtained using 5000 independent samples from each distribution.}}
\label{fig:DP4}
\end{figure}

\begin{figure}[h]
\centering
\includegraphics[width = 0.4\columnwidth]{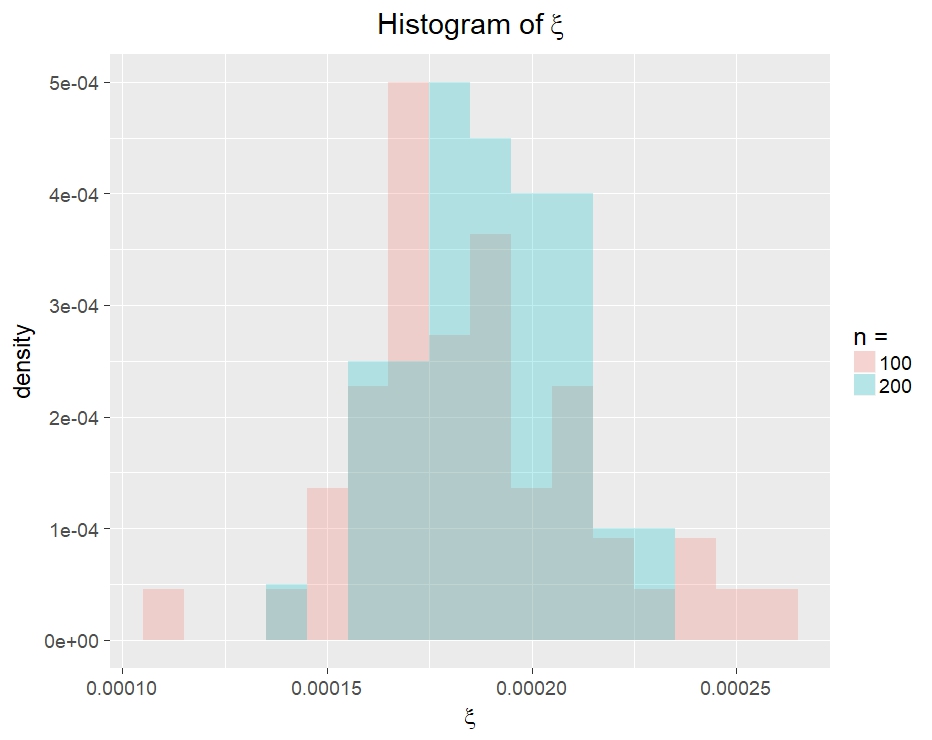}
\includegraphics[width = 0.4\columnwidth]{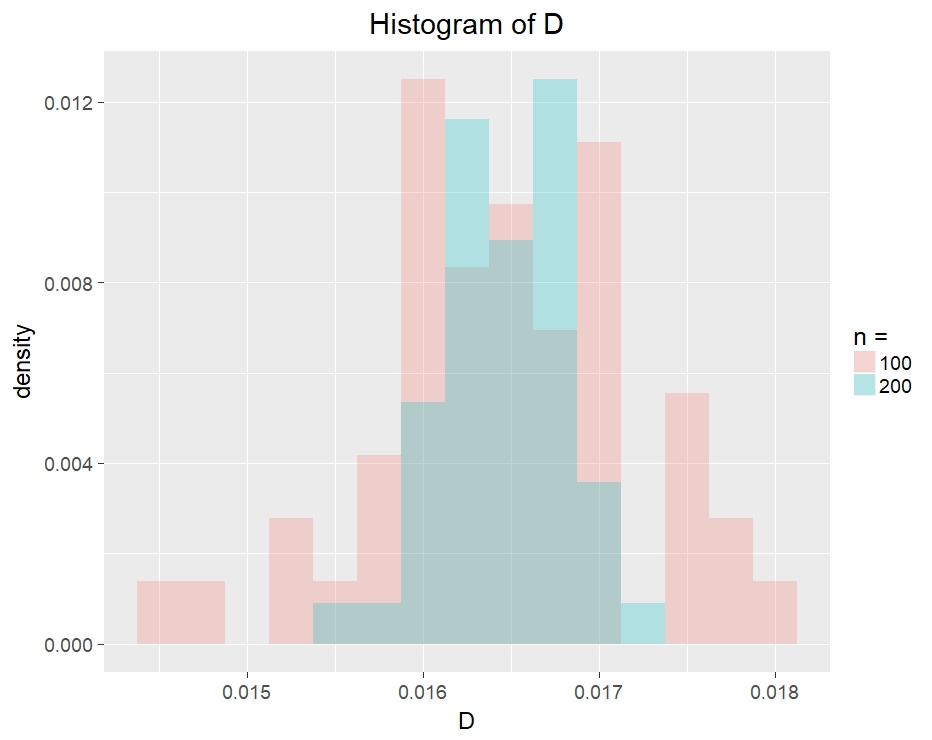} 
\caption{{\em Histogram of $\xi$ (left panel) and $D$ (right panel) over 50 independent replicates for the Probit-Gaussian Process simulation. The pink is when $n = 100$ and the blue is when $n = 200$.}}
\label{fig:ksi2}
\end{figure}

Figures \ref{fig:DP3} and \ref{fig:DP4} show the marginal density plots of 8 coordinates of $\theta$ based on the 5000 samples for the two values of $n$ respectively. The tMVN distribution is shown in blue while the soft tMVN is in pink. It is evident that for both values of $n$, the marginal densities are visually indistinguishable. To obtain an overall summary measure, Figure \ref{fig:ksi2} shows the histogram of $\xi$, defined in equation \eqref{eq:xi}, (left panel) and $D$, defined in \eqref{eq:D}, over $50$ independent simulations. Both the histograms are tightly centered near the origin, which again suggests the closeness of the tMVN and soft tMVN distributions. As a quick comparison, the value of $D$ between $\Gauss(0, \Sigma)$ and $\Gauss(0.005, \Sigma)$  for the current $\Sigma$ is about $0.03$ for both values of $n$.

\subsection{Probit-Gaussian Example}\label{sect:probit}

Our second example assumes $\theta \sim \m N_{N+P}(0, \Sigma) \ind_{\m{C}}(\theta)$ where
$$\Sigma = \begin{bmatrix} I_n + X \Lambda X^{\T} & X \Lambda \\ \Lambda X^{\T} & \Lambda \end{bmatrix},$$
$\m{C} = \m{C}_1 \otimes \m{C}_2 \otimes \cdots \otimes \m{C}_N \otimes \bbR^P$, $\m{C}_i$ is either $(-\infty, 0)$ or $(0, \infty)$ for $i$ in $1, \ldots, N$, $X$ is an $N \times P$ matrix, and $\Lambda$ is a $P \times P$ diagonal matrix.

This covariance structure is motivated by a univariate/multivariate probit model. The usual univariate probit model has binary response variables $Y_i = \{0,1\}$ with predictors $x_i \in \bbR^d$ for $i = 1, \ldots, n$. Using the latent variable representation of \cite{albert}, $Y_i = \ind(z_i > 0)$ where $z_i$ follows a $\m N(x_i^{\T}\beta, 1)$ distribution and $\beta \in \bbR^p$. Setting a Gaussian prior on $\beta$, $\beta_j \sim \m N(0, \lambda_j)$, the joint distribution of $\theta = [z, \beta]$ follows a Gaussian distribution. Then the conditional posterior of $\theta \mid Y, x, \lambda$ follows the above $\m N_{N+P}(0, \Sigma)\ind_{\m{C}}(\theta)$ distribution where $X = [x_1, \ldots x_n]^{\T}$, $\Lambda = \diag\{\lambda_1, \ldots, \lambda_p\}$, $N = n$, $P = p$, and $\m{C}_i = (-\infty, 0)$ if $Y_i = 0$ and $\m{C}_i = (0, \infty)$ if $Y_i = 1$. 

The multivariate probit model has data $(y_i, x_i)$ where $y_i = [y_{i1}, \ldots, y_{iq}] \in \{0,1\}^q$ is a binary response with predictors $x_i \in \bbR^p$ for $i = 1, \ldots n$. Using data augmentation, $y_{ik} = \ind(z_{ik})$ where $z_{ik}$ follows a $\m N(x_i^{\T}\beta_k,1)$ distribution and $\beta_k \in \bbR^p$. Assume that $\beta_{jk}$ follows a $\m N(0, \lambda_{jk})$ prior. Letting $\tilde{y}_k = [y_{1k}, \ldots, y_{nk}]$, $\tilde{z}_k = [z_{1k}, \ldots, z_{nk}]$, and $\lambda_k = [\lambda_{1k}, \ldots, \lambda_{pk}]$, we can rewrite the model in terms of vectors instead of matrices. Let $Y = [\tilde{y}_1, \ldots, \tilde{y}_q]$, $Z = [\tilde{z}_1, \ldots, \tilde{z}_q]$, $\lambda = [\lambda_1, \ldots, \lambda_q]$, and $\beta = [\beta_1, \ldots, \beta_q]$. Then $\theta = [Z, \beta]$ follows a Gaussian distribution and the conditional distribution of $\theta$ follows the above $\m N_{N+P}(0, \Sigma)\ind_{\m{C}}(\theta)$ where $\tilde{X} = [x_1, \ldots, x_n]^{\T}$, $X = \diag(\tilde{X})_{k = 1, \ldots, q}$, $\Lambda = \diag(\lambda)$, $N = nq$, $P = pq$, and $\m{C}_{ik} = (-\infty, 0)$ if $y_{ik} = 0$ and $\m{C}_{ik} = (0, \infty)$ if $y_{ik} = 1$.

For this simulation, we sample $x_i \stackrel{iid}{\sim} \m N(0,I_P)$ and $\lambda_j \sim U[1/15, 1/5]$, and then set $\Sigma$ to the above form. Draw $\beta \sim \m N(0, \Lambda)$ and $Z \sim \m N(X\beta, I_n)$. Then if $Z_i \geq 0$, set $Y_i = 1$ and if $Z_i < 0$, set $Y_i = 0$. For both $(N, P) = \{(100, 400), (200, 400)\}$, we then proceed to draw 5000 samples from the tMVN, $\m N_{n}(0, \Sigma)\ind_{\m{C}}(\theta)$, using Botev's rejection sampler and 5000 samples from the soft tMVN, $\m N_{n}^s(0, \Sigma)\ind_{\m{C}}(\theta)$, using our Gibbs sampler. The 5000 samples were collected for our method after discarding 1000 initial samples as burn-in and collecting every 100th sample to thin the chain. 
 
\begin{figure}[htp!]
\centering
\includegraphics[width = 1.1\columnwidth]{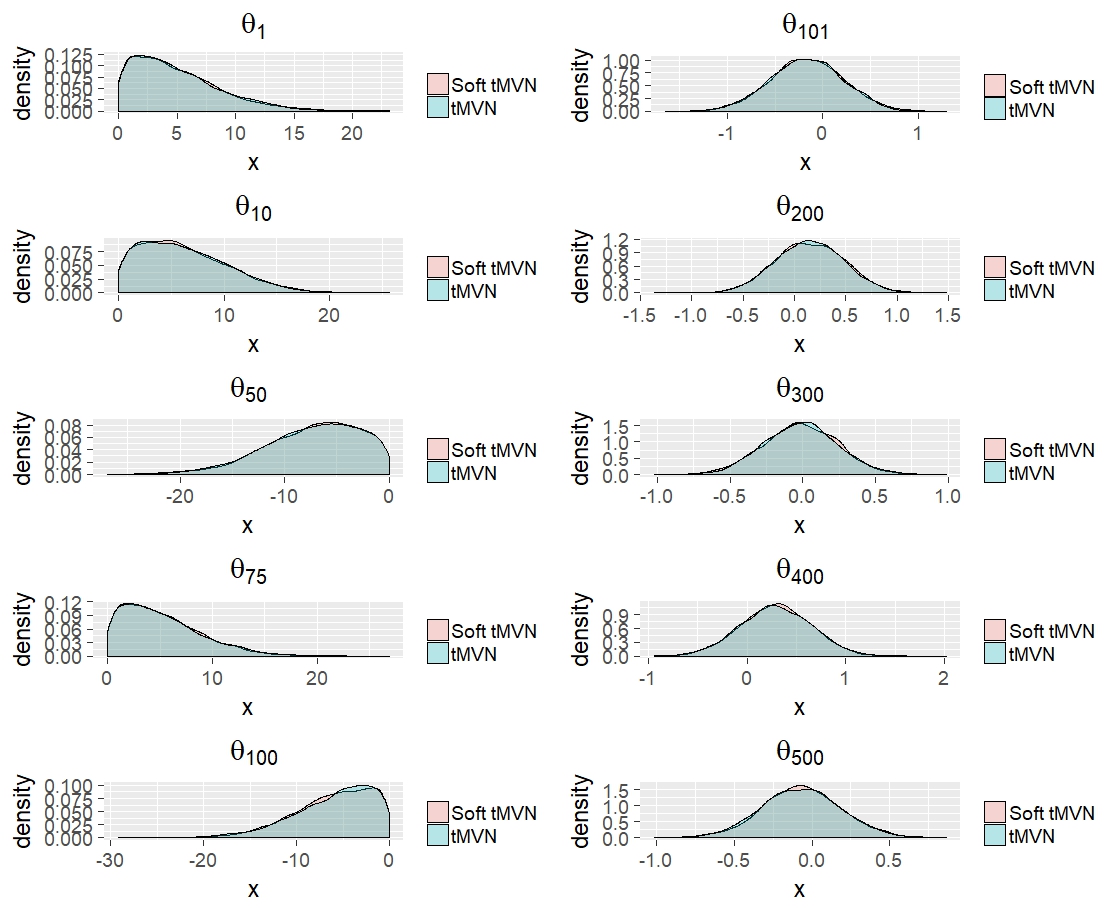}
\caption{{\em Overlapping density plot for the Probit-Gaussian simulation when $n = 100$. Blue denotes tMVN and pink denotes the soft tMVN distribution. The density plots are obtained using 5000 independent samples from each distribution.}}
\label{fig:DP1}
\end{figure}

\begin{figure}[htp!]
\centering
\includegraphics[width = 1.1\columnwidth]{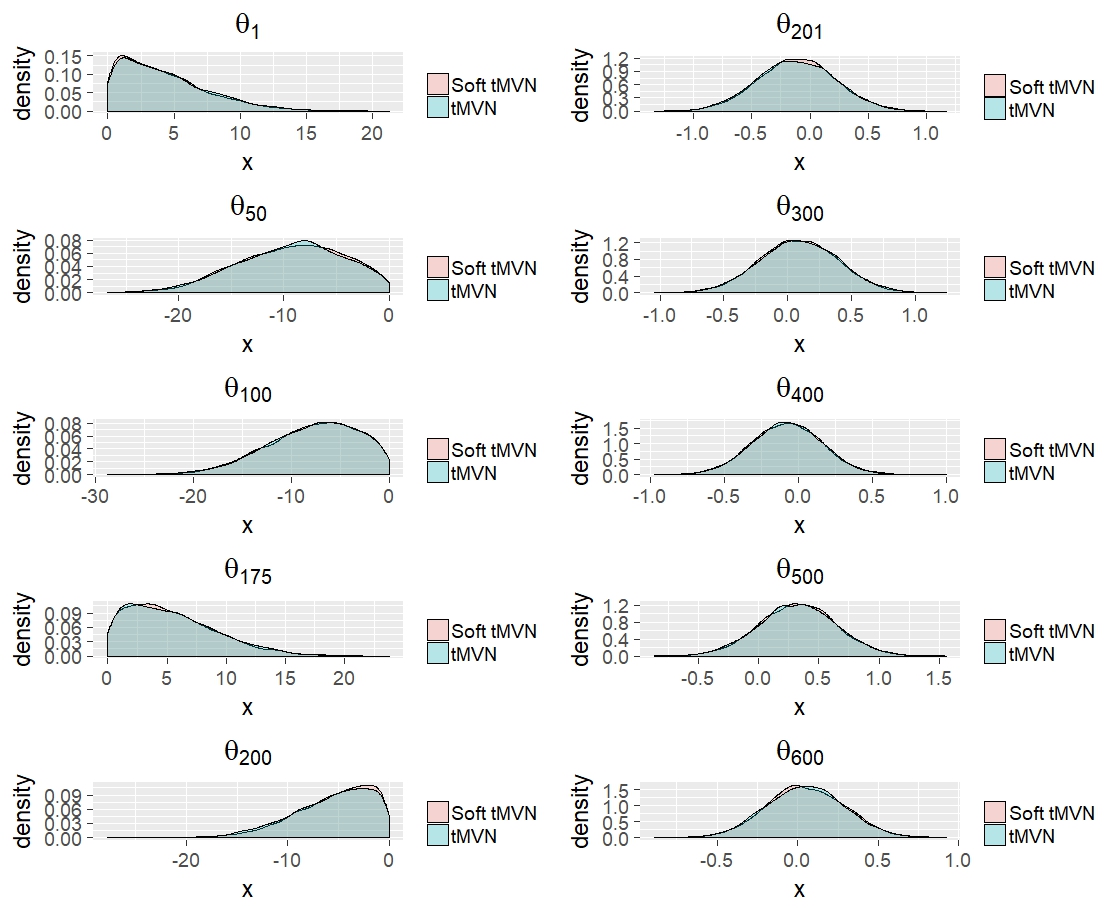}
\caption{{\em Overlapping density plot for the Probit-Gaussian simulation when $n = 200$. Blue denotes tMVN and pink denotes the soft tMVN distribution. The density plots are obtained using 5000 independent samples from each distribution.}}
\label{fig:DP2}
\end{figure}

\begin{figure}[htp!]
\centering
\includegraphics[width = 0.4\columnwidth]{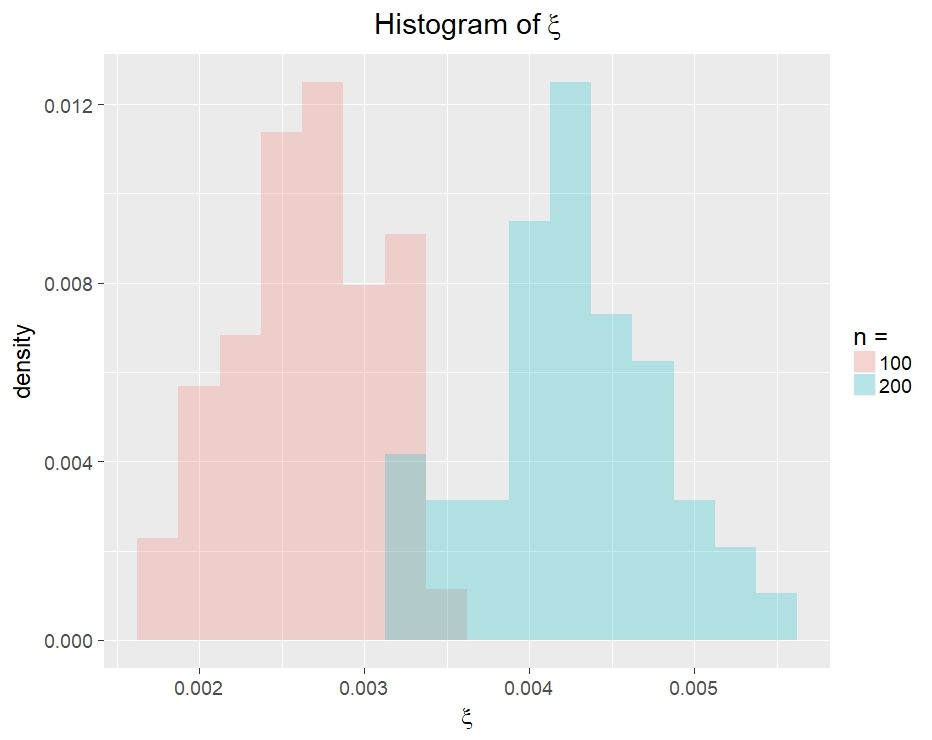} 
\includegraphics[width = 0.4\columnwidth]{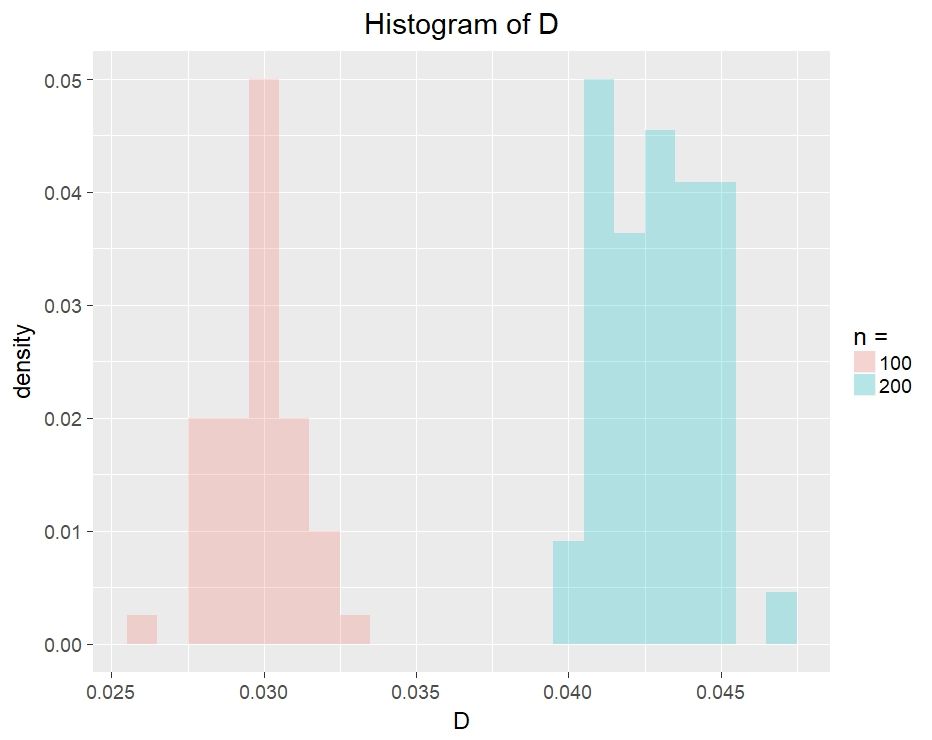}
\caption{{\em Histogram of $\xi$ (left) and $D$ (right) over 50 trials for the Probit-Gaussian simulation. The pink is when $n = 100$ and the blue is when $n = 200$.}}\label{fig:ksi1}
\end{figure}

\begin{figure}[htp!]
\centering
\includegraphics[width = 0.5\columnwidth]{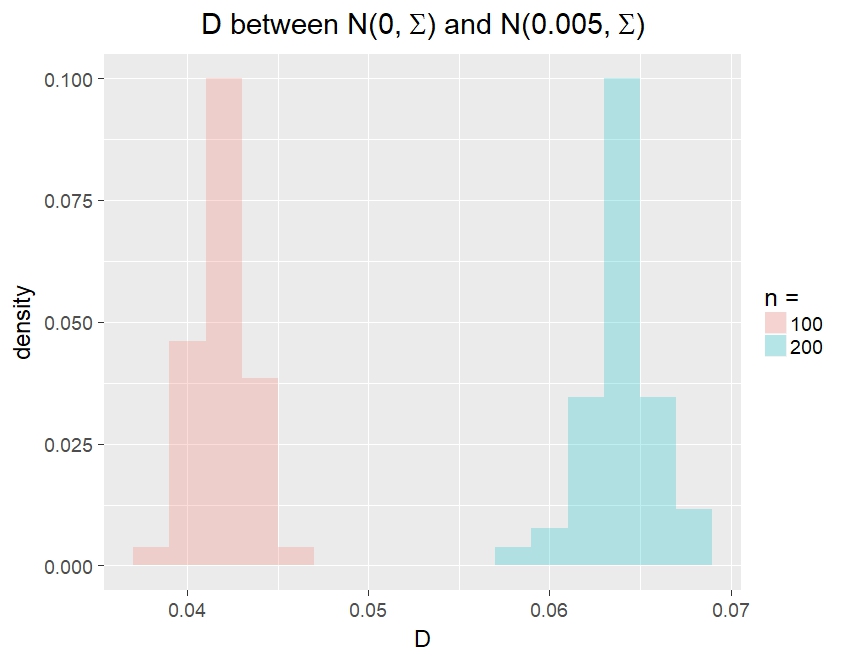} 
\caption{{\em Histogram of $D$ over 50 trials between $\m N(0, \Sigma)$ and $\m N(0.005, \Sigma)$ where $\Sigma$ is the same as in the Probit-Gaussian simulations. The pink is when $n = 100$ and the blue is when $n = 200$. This is used for comparison with Figure \ref{fig:ksi1}.}}\label{fig:wass_cal}
\end{figure}

Figures \ref{fig:DP1} and \ref{fig:DP2} show the marginal density plots of 8 coordinates of $\theta$ based on the 5000 samples for the two combinations respectively; as before, the tMVN distribution is shown in blue while the soft tMVN is in pink. We once again see that for both combinations, the marginal densities overlap well. To obtain an overall summary measure, Figure \ref{fig:ksi1} shows the histogram of $\xi$, defined in equation \eqref{eq:xi}, (left panel) and $D$, defined in \eqref{eq:D}, over $50$ independent simulations. We see that the histogram of $\xi$ and $D$ shifts to the right for $n = 200$ than for $n = 100$. This shift is expected as the size of the matrix $X$ grows, and thus, the size of $\Sigma$ grows. As a point of comparison, in Figure \ref{fig:wass_cal}, we plot the histogram of $D$ between $\m N(0, \Sigma)$ and $\m N(0.005, \Sigma)$  for the present choice of $\Sigma$ and see a similar shift. We believe that the shift occurs for the probit-Gaussian motivated soft tMVN but not the probit-Gaussian process motivated soft tMVN due to structure of $\Sigma$. In the probit-Gaussian process motivated soft tMVN, $\Sigma$ does not change with each trial and it has a very solid structure, while in the probit-Gaussian motivated soft tMVN, $\Sigma$ changes for each trial and has a very random structure.

\section{Usage as prior in Bayesian constrained regression}\label{sec:cons}
In this section, we provide a concrete example of using the soft tMVN distribution as a prior distribution in a constrained Gaussian regression problem. As noted in the introduction, a general approach to Bayesian constrained regression is to expand the unknown function onto a suitable basis which allows formulation of the functional constraints in terms of linear constraints on the basis coefficients. Since the soft tMVN distribution is also conditionally conjugate to a Gaussian likelihood, one may use it as a prior distribution on the basis coefficients instead of a tMVN distribution. For illustration purpose, we consider a monotone single-index model considering its usefulness in practical applications, noting that the methodology can be extended to more standard constrained regression applications such as estimation of bounded, monotone, or convex/concave functions. We pick the monotone single-index model example due to limited previous treatment from a Bayesian perspective. Moreover, this example nicely brings out the computational advantages of using a soft tMVN prior. 

Given response-covariate pairs $\{(y_i, x_i)\}_{i=1}^n \in \bbR \times \bbR^p$, a Gaussian single index model \citep{Antoniadis,Chen,Gramacy,Wang2009,Yu2002} assumes the form 
\begin{align}\label{eq:sing_mod}
y_i = f(x_i^{\T}\alpha) + \epsilon_i, \quad \epsilon_i \sim N(0, \sigma^2), \quad i = 1, \ldots, n,
\end{align}
where $f : \bbR \to \bbR$ is an unknown link function and $\alpha \in \bbR^p$ an unknown coefficient vector. Throughout, we assume the covariates to be standardized. The single-index model provides a bridge between linear and non-linear modeling  by first linearly projecting the high-dimensional vector of predictors to the real line and then modeling the response as a non-linear function of the projection. The model \eqref{eq:sing_mod} is clearly non-identifiable without further restrictions; we follow a standard prescription to impose a unit norm restriction, $\|\alpha\| = 1$, on $\alpha$. 

We consider a monotone single-index model \citep{Cavanagh,Ahn, Balabdaoui,Foster,Luo} where the link function $f$ is monotone non-decreasing. Monotone single-index models have widespread applications in biomedical science, e.g.  find gene-gene interactions \citep{Luss} and to study the relationship between risk factors for survival with leukemia \citep{Schell}. To model $f$, we use a Bernstein polynomial basis noting that other basis functions mentioned in the introduction can also be used. 
Using the Bernstein polynomial basis, there are established sufficient conditions which enforce $f$ to be monotonic. Define, for $j = 0, \ldots, M$, 
$$B_{M,j}(u) = \binom{M}{j}u^j(1-u)^{M-j}, \quad u \in [0, 1],$$
so that the Bernstein polynomial of degree $M$ is
$$B_M(u) = \sum_{j=0}^M \theta_j B_{M,j}(u).$$
If 
\begin{equation}\label{eq:non-decreasing}
\theta_0 \leq \theta_1 \leq \cdots \leq \theta_M,
\end{equation}
then $B_M(u)$ is non-decreasing \citep{Chak}. 

To apply the Bernstein polynomial basis to our setting, we need some preprocessing as described below. Since $|x_i^{\T}\alpha| \leq \norm{x_i}\norm{\alpha} = \|x_i\|$ by the Cauchy-Schwarz inequality and the identifiability restriction respectively, 
if we let $c = \max_{i} \norm{x_i}$ and transform $\tilde{x_i} = x_i/c$, we have $|\tilde{x_i}^{\T}\alpha| \leq 1$. Hence, we need to perform a change of variable to transform the support of the Bernstein polynomial to $[-1,1]$. To that end, we write $B_{M,j}(u) = p_j(u)/(M+1)$ for $u \in [0, 1]$, where $p_j(u)$ is the density of a Beta$(j+1, M-j+1)$ distribution. Letting $T = 2U-1$ for $U \sim \mbox{Beta}(j+1, M-j+1)$, the density of $T$ is $q_j(t) = \frac{1}{2}p_j\{(t+1)/2\}$ for $t \in [-1,1]$. Let $\tilde{B}_{M,j}(t) = q_j(t)/(M+1)$ for $j = 0, \ldots, M$ represent the transformed Bernstein polynomial basis and define our monotone single-index model as 
\begin{align}\label{eq:mon_sing_mod}
y_i = \tilde{B}_M(\tilde{x}_i^{\T}\alpha) + \epsilon_i, \quad \tilde{B}_M(t) = \sum_{j=0}^M \theta_j \tilde{B}_{M,j}(t), \quad t \in [-1, 1].
\end{align}
Under the order-restriction on the basis coefficients in \eqref{eq:non-decreasing}, $\tilde{B}_M(\cdot)$ remains non-decreasing. Set $\psi_0 = \theta_0$, $\psi_1 = \theta_1-\theta_0, \ldots, \psi_M = \theta_M - \theta_{M-1}$, so that \eqref{eq:non-decreasing} is equivalent to $\psi_k \geq 0$ for $k = 1,\ldots,M$. Thus the non-decreasing constraint can be written in terms of $\psi= [\psi_0, \ldots, \psi_M]^{\T}$. Let $A$ be an $(M+1) \times (M+1)$ lower triangular matrix where all the lower triangle elements and diagonal elements are 1. Then $A\psi = \theta$ where $\theta = [\theta_0, \ldots, \theta_M]^{\T}$.

To place the monotone single-index model \eqref{eq:mon_sing_mod} in vectorized notation, let $\tilde{B}_M^i = [\tilde{B}_{M,0}(\tilde{x}_i^{\T}\alpha),\\ \ldots, \tilde{B}_{M,M}(\tilde{x}_i^{\T}\alpha)]^{\T}$ and $\mathbb{B}_{\alpha} = [\tilde{B}_M^1, \ldots \tilde{B}_M^n]^{\T}$ so that $\mathbb{B}_\alpha$ is a $n \times (M+1)$ matrix, with the subscript serving as a reminder that $\mathbb{B}_{\alpha}$ depends on $\alpha$. Then letting $Y = [y_1, \ldots, y_n]^{\T}$, \eqref{eq:mon_sing_mod} can be equivalently represented as
$$Y = \mathbb{B}_\alpha \theta + \epsilon = \mathbb{B}_\alpha A \psi + \epsilon.$$

Our prior specification on the model parameters $(\psi, \alpha, \sigma^2)$ assumes the form $\pi(\psi, \alpha, \sigma^2) = \pi(\psi) \, \pi(\alpha) \, \pi(\sigma^2)$. We consider two different priors on $\psi$: (i) a tMVN prior $\m N_{\m C}(0, 25 I_{M+1})$, and (ii) a soft tMVN prior $\m N_{\m C}^s(0, 25 I_{M+1})$, where in both cases $\m C = \bbR \otimes [0, \infty)^M$. Next, we set $\alpha = \beta/\norm{\beta}$ and assign a standard Gaussian prior on $\beta$. Finally, we consider a inverse-Gamma prior on $\sigma^2$ with mean 1 and variance 10. For sake of future reference, we refer to the joint prior on $(\psi, \alpha, \sigma^2)$ corresponding to cases (i) and (ii) by $\pi^h$ and $\pi^s$ respectively, with the superscripts indicative of a usual (hard) or soft tMVN prior on the constrained parameter. 

We employ a Metropolis-within-Gibbs algorithm to sample from the posterior distribution with either prior. For $\pi^h$, the conditional posterior $\psi \mid \sigma^2, \alpha$ is $\m N_{\m C}(\mu_\psi, \Sigma_\psi)$, while the same for $\pi^s$ is $\m N_{\m C}^s(\mu_\psi, \Sigma_\psi)$, where 
$$\Sigma_\psi = \left(\frac{1}{\sigma^2}D_{\alpha}^{\T}D_\alpha + \frac{1}{25}I_{M+1}\right)^{-1}, \quad \mu_{\psi} = \frac{1}{\sigma^2}\Sigma_\psi D_\alpha^{\T} Y, \quad D_\alpha = \mathbb{B}_\alpha A.$$
The conditional distribution of $\sigma^2|\psi,\alpha$ is inverse-Gamma in both cases. To sample from $\alpha|\sigma^2,\psi$,  we use a Metropolis step with the proposal density on $\beta$ as $J(\beta^t|\beta^{t-1}) \sim \m N(\beta_{t-1},0.01^2I)$. The proposal standard deviation of $0.01$ was chosen to give an acceptance probability around 0.35 for $\beta$. 

The following simulation compares the Metropolis-within-Gibbs algorithms for the priors $\pi^h$ and $\pi^s$ respectively. 
We generate data from the model \eqref{eq:mon_sing_mod} with $n = 800$, $p = 5$, $M = 20$, and a set of true parameter values $\psi_0, \alpha_0, \sigma_0^2$. We set $\sigma_0 = 0.1$ and $\alpha_0 = \beta_0/\|\beta_0\|$ with $\beta_0$ drawn from a standard Gaussian distribution. Finally, we set $\theta_0 \in \mathbb{R}^{21}$ equal to the vector where the first six entries are -1, then -0.5, then the next seven entries are 0, then 0.5, then the last six entries are 1. We consider $30$ independent replicates for model fitting and perform out-of-sample prediction on a single separate dataset of size $200$. 

We set $\eta = 500$ for the soft tMVN prior $\pi_s$. We observed sensitivity for smaller values of $\eta$ in this context; something that we didn't encounter earlier, possibly due to the more difficult sampling problem involved here\footnote{See the supplemental document for an example with a smaller value of $\eta$.}.
For each of the 30 replicates, we run the Gibbs samplers for $\pi_h$ and $\pi_s$ outlined above to collect 1000 posterior samples each. These 1000 samples are after a burn-in period of 1000 and after thinning the chain by 100. The 1000 samples are used to calculate the posterior mean of $\alpha$, $\hat{\alpha}$, and the posterior mean of $\theta$, $\hat{\theta}$.
For $\pi_h$, we use the rejection sampler of \cite{botev} implemented in the \texttt{R} package \texttt{TruncatedNormal} \citep{truncatednormal} to draw samples from the tMVN distribution, while for $\pi_s$, we use our data augmentation Gibbs sampler to sample from the soft tMVN distribution. The code to run both Gibbs samplers can be found at \url{https://github.com/aesouris/softTMVN}.

In terms of statistical performance, the two samplers were comparable. The average out-of-sample prediction error for the soft tMVN prior across the 30 replicates was $0.005$ with a standard deviation of $0.0106$, while the same numbers for the tMVN prior were $0.002$ and $0.0066$ respectively. 

\begin{table}[h]
{\centering
\begin{tabular}{r|ccc}
                & $\alpha$-ESS          & $\psi$-ESS     & run-time (in hours)       \\ \hline
soft tMVN prior & $253.6625$ & $686.0742$ & $3.78_{0.0026}$ \\
tMVN prior      & $168.4741$ & $796.6799$ & $15.45_{3.8204}$
\end{tabular}
\caption{ {\em The first two columns report the average effective sample sizes (out of 1000 MCMC samples) for $\alpha$ and $\psi$ for the two Gibbs samplers. The average is over both the parameter entries as well as the 30 replicates. The final column reports the run-time (in hours) for the respective Gibbs samplers to collect 1000 posterior samples, with the subscript denoting the standard deviation across replicates.}} \label{tab:eff} 
}
\end{table}

Table \ref{tab:eff} reports the effective sample sizes for $\alpha$ and $\psi$ as well as the run-time for the two Gibbs samplers. The two samplers are similar in terms of the effective sample sizes; however the Gibbs sampler for the tMVN prior has almost 5 times the run-time of the soft tMVN sampler. The mixing is slow for either samplers which is indicative of a general issue for problems with constrained parameter spaces; remember the 1000 posterior samples are collected with a thinning size of 100. Although a formal proof is beyond the scope of the paper, empirical evidence suggests that the constrained parameters inside the Gibbs sampler may get stuck into regions of low probability, and it can take a long time to escape these regions. Specifically, we see that Botev's state-of-the-art rejection sampler can sometimes take exceedingly long to make a single move; note the variability in the run-time across the 30 trials in Table \ref{tab:eff}.  While our chain also suffers from a similar slow mixing, it has substantially better per-iteration cost which makes it possible to run it for a large path-length to collect a substantial number of effective samples. The computational advantage becomes even more pronounced for higher dimensions; we do not report a simulation with a higher dimension $M$ since the tMVN sampler takes exceedingly long to run. 

\section{Discussion}
In this paper, we have presented the soft tMVN distribution, which provides a smooth approximation to the tMVN distribution with linear constraints. Our theoretical and empirical results suggest that the soft tMVN distribution offers a good approximation to the tMVN distribution in high dimensional situations. We envision the soft tMVN distribution to be applicable in Bayesian constrained problems as a more computationally viable alternative prior to the usual tMVN prior, especially in complex problems where the an MCMC algorithm may get stuck in regions of very low probability under a tMVN prior, making it difficult to move. The monotone single index model example illustrates this phenomenon and we expect it to be more widely prevalent. 

\clearpage
\bibliographystyle{humannat}
\bibliography{soft_tMVN}

\end{document}